\font\small=cmr10 scaled 1000         
%
%
\magnification=\magstep1
\baselineskip=11pt plus .1pt minus .1pt
\hsize=12.5truecm
\vsize=19.0truecm  
\hfuzz=5pt\vfuzz=5pt
\tolerance=1000
\overfullrule=0pt
\parskip=0pt
\abovedisplayskip=3 mm plus6pt minus 4pt
\belowdisplayskip=3 mm plus6pt minus 4pt
\abovedisplayshortskip=0mm plus6pt minus 2pt
\belowdisplayshortskip=2 mm plus4pt minus 4pt
\predisplaypenalty=0
\clubpenalty=10000
\widowpenalty=10000
\parindent=2em
%
%
\font\pgnumfont=cmr9
\font\headlinefont=cmti9
\font\titlefont=cmbx10
\font\authorfont=cmr10
\font\addressfont=cmti9
\font\datefont=cmr9
\font\sumfont=cmr9

\font\absfont=cmbx9
\font\secfont=cmr10
\font\subsecfont=cmti10
\font\subsubsecfont=cmr10
\font\figfont=cmr9
\font\figheadfont=cmbx9

\font\tabheadfont=cmbx9
\font\mainfont=cmr10
\font\petitrm=cmr9

%
%
%
\newtoks\TITLE \newtoks\AUTHOR \newtoks\ADDRESS \newtoks\SUMMARY
\newdimen\sumindent \sumindent=\parindent
\newtoks\KEYWORDS \newtoks\SUBMITTED \newtoks\ACCEPTED
\newtoks\SENDOFF
%

%
%
\newtoks\firstpage
\let\firstpage=Y
\newtoks\AUTHORHEAD \newtoks\ARTHEAD \newtoks\VOLUME \newtoks\PAGES
\if!\the\AUTHORHEAD!\AUTHORHEAD={\the\AUTHOR}\fi
\if!\the\ARTHEAD!\ARTHEAD={\the\TITLE}\fi
\footline={\hfil}
\headline={\ifodd\pageno\rightheadline \else\leftheadline\fi}
\def\leftheadline{\if Y\firstpage\firsthead\global\let\firstpage=N
  \else\lefthead\fi}
\def\rightheadline{\if Y\firstpage\firsthead\global\let\firstpage=N
  \else\righthead\fi}
\def\lefthead{\pgnumfont\number\pageno\hfil\headlinefont\the\AUTHORHEAD}
\def\righthead{\headlinefont\the\ARTHEAD\hfil\pgnumfont\number\pageno}
\def\firsthead{\headlinefont Baltic Astronomy,~vol.\the\VOLUME,
\the\PAGES,~\the\year .\hfil}
\voffset=2\baselineskip 
%

\newdimen\oldbaselineskip \oldbaselineskip=\baselineskip
\def\test#1{\newlinechar=`@\if!\the#1! \message{#1 not given@}\fi}%
\def\printheader{
  \parindent=0pt
  \null\vskip1.cm
  \test{\TITLE}
  \vbox{\baselineskip=15pt
    \titlefont\the\TITLE
    }
  \vskip8mm plus8mm
  \test{\AUTHOR}
  \authorfont\the\AUTHOR
  \vskip2mm
  \test{\ADDRESS}
  \addressfont\the\ADDRESS
  \vskip2mm
  \test{\SUBMITTED}
  \line{\datefont Received \the\SUBMITTED
    \if!\the\ACCEPTED!\else, accepted \the\ACCEPTED\fi.\hfill}
  \vskip4mm plus4mm
  \vbox{\leftskip=\sumindent\parindent=0pt
    \parskip=5pt
    \absfont Abstract.
    \test{\SUMMARY}
    \sumfont\the\SUMMARY\par
    \absfont Key words:
    \test{\KEYWORDS}
    \sumfont\the\KEYWORDS\par
    }
  \sumfont
  \if!\the\SENDOFF!\else\footnote{}{
 \the\SENDOFF}\fi
  \parindent=2em
  }
%
%
\newdimen\uppergap \newdimen\lowergap
\uppergap=5mm \lowergap=3mm
\newdimen\secind \newdimen\subsecind \newdimen\subsubsecind
\setbox0=\hbox{\secfont 9. }\secind=\wd0
\setbox0=\hbox{\subsecfont 9.9. }\subsecind=\wd0
\setbox0=\hbox{\subsubsecfont 9.9.9. }\subsubsecind=\wd0
\def\section#1{\goodbreak\par\vskip\uppergap
  \noindent\hangindent\secind\hangafter=1\secfont#1
  \vskip\lowergap\mainfont\par\nobreak}
\def\subsection#1{\goodbreak\par\vskip\uppergap
  \noindent\hangindent\subsecind\hangafter=1\subsecfont#1
  \vskip\lowergap\mainfont\par\nobreak}
\def\subsubsection#1{\goodbreak\par\vskip\uppergap
  \noindent\hangindent\subsubsecind\hangafter=1\subsubsecfont#1
  \vskip\lowergap\mainfont\par\nobreak}
%
%
%

%

%
\newdimen\tabind
\setbox0=\hbox{\tabheadfont Table 55.} \tabind=\wd0

%
%
\def\References{\vskip\uppergap
\line{\secfont REFERENCES\hfill}
  \vskip0.8\lowergap
 \petitrm
  }
\def\ref{\goodbreak
\hangindent12pt\hangafter=1
\noindent\ignorespaces}
\def\endref{\egroup}

\def\ref{\goodbreak
\hangindent12pt\hangafter=1
\noindent\ignorespaces}
\def\endref{\egroup}
%
%
\def\byebye{\egroup\par\vfill\supereject\end}
%
%

%
%

\def\utw{\smash{\rlap{\lower5pt\hbox{$\sim$}}}}
\def\udtw{\smash{\rlap{\lower6pt\hbox{$\approx$}}}}

\newdimen\free\newdimen\shift
\def\Entry#1#2#3{\par\goodbreak\smallskip%
  \setbox1=\vbox{\advance\hsize by-10mm\parindent=0pt
    \def\\{\par}%
    \it#1. \rm#2}
  \line{\box1\hfill#3}\smallskip
}%
\newdimen\savesize

\def\shiftfigure #1#2#3#4#5{
    \vbox to #2 { \ifodd #5 \rightskip#4 \else\leftskip#4 \fi
                  \null\vfil
                  \figheadfont Fig.~#1.\figfont #3
                  \medskip
                }
                          }

\year1996

\VOLUME{0}
\PAGES{00-00}
\pageno=1

\vskip -1.0 cm               
\noindent  
{\small  
Proc.~~{\it Internat.~Cooperation in Dissemination of Astronomical Data},} \break
\vskip -7 mm
{\small   
\centerline{July 3--5, 1996,~~St.-Petersburg, Russia}} 
\vskip -0.6 cm

\TITLE={PREPARING A PUBLIC DATABASE OF RADIO SOURCES}
\AUTHOR={H.~Andernach$^1$, S.A.~Trushkin$^2$, A.G.~Gubanov$^3$,
O.V.~Verkhodanov$^2$, V.B.~Titov$^3$ and A.~Micol$^4$}

\ARTHEAD={Public Database of Radio Sources} 
\AUTHORHEAD={H.~Andernach et al.}    

\ADDRESS={$^1$IUE Observatory, Villafranca, Apdo.~50727, E--28080 Madrid, Spain

$^2$Special Astrophysical Observatory RAS, Nizhnij Arkhyz, 357147 Russia

$^3$Astronomical Inst., St.-Petersburg State University, 198904 Russia

$^4$ST-ECF, Karl-Schwarzschild-Str.~2, D--85748 Garching, Germany}

\SUBMITTED={July 20, 1996}

\SUMMARY={
We have collected the largest existing set of radio source 
lists in machine-readable form~: 320 tables with 1.75 million 
records.  Only a minor fraction of these is accessible 
via public databases. We describe our plans to make this huge
amount of heterogeneous data accessible in a homogeneous way 
via the World Wide Web, with reliable cross-identifications, 
and searchable by various observables. 
         }

\KEYWORDS={
radio sources: general; databases; catalogues
          }

\printheader

\vskip 5 mm

Surveys of radio sources have always been instrumental in widening 
our horizon of the distant Universe:~~optical identifications of the
sources are pointing us to the most distant galaxies known,
and the perfect isotropy of the source distribution in even the deepest 
surveys shows that we are sampling a {\it much} larger volume of space
than we do with magnitude-limited galaxy surveys.
Of course, an efficient search for a certain type of objects (e.g. distant
galaxies, variable sources, etc.) requires the
filtering of large samples of radio sources carefully selected from
literature by very specific criteria.
Until now this required to construct these various 
samples by tedious merging and sorting of numerous individual source
catalogues. In addition, existing astronomical databases 
show a notable lack in published information on radio sources,
and data centers provide only a small part of the largest published
source catalogues.

One of us (Andernach 1990) has noted this lack years ago.
Since then he has gathered the yet most complete set of radio
source catalogues in electronic form~: 320 source tables with 
a total of 1.75 million entries. Only a minor fraction of the tables is
accessible through public databases.
This collection is the result of six years of persistent requests to the 
authors, even though in an early email campaign radio astronomers were
invited to contribute their electronic source tables to the 
astronomical data centres voluntarily. 
In an effort to recover also the largest published source lists {\it not}
available in electronic form, H.A. employed a scanner and 
``Optical Character Recognition'' (OCR) software. Since early 1995
over 100 printed data tables with $>$50,000  entries into electronic
form. Staff at Special Astrophysical Observatory of the Russian 
Academy of Sciences (SAO RAS), also helped correcting, editing and 
proof-reading of the raw OCR result, as well as in the manual retyping of 
many source tables which were unrecoverable with OCR. 
Tables published {\it only} on microfiche are the most difficult ones
to recover, since microfiche copiers have turned
into an extremely rare species of apparatus.

Independently, SAO RAS and its St.-Peters\-burg branch (SPbB SAO) and the
Astronomical Institute of the St.-Peters\-burg State University (AI SPbU)
maintain a collection of astronomical catalogues. The database of ``Radio
Astronomical Catalogues'' (RAC) has been created at AI (Gubanov \& Titov
1996;  http://www.aispbu.spb.su/WWW/RAC.html).
The ``CATalogue support System'' (CATS) is
being developed at SAO (Verkhodanov \& Trushkin 1995; 
telnet {\tt ratan.sao.ru}, login as {\tt cats}, pw=$<$CR$>$). 
Both are supported by the Russian Foundation for Basic Research (RFBR).

A. Micol is involved in the use of relational databases 
for astronomical applications.  He will be responsible for the
design of the database structure and its user interface,
translation of user to software requirements,
configuration and maintenance of a WWW server,
and for application Program Interfaces (API) to commercial
database management systems.

The present project joins the efforts of these groups
to establish the first reasonably complete and publicly accessible
database of radio sources.
During the first half year of active collaboration 
over fifty tables and catalogues were either scanned or typed,
proof-read and included in the full collection of data sets.
Tools were developed to cross-identify sources from different
catalogues and to construct and display their radio spectra.

Virtually all catalogues have a different format and list different 
observables.   It will be a major challenge to provide
uniform access to such a heterogeneous collection of data sets involving
different methods, notations and units.
Only a clear understanding of observing techniques and data-processing 
for a given catalogue will guarantee reliable cross-IDs between
catalogues. Thus, documentation files in an agreed format will be prepared 
for all catalogues accessible to the search.
Many catalogues provide only source names and will have to be completed
with positional data. For other tables lacking positional errors we will
insert these errors from formulae given in the publication.

The concept of a ``Reference Directory'' (a central repository of
metadata like the descriptions of catalogue fields, their physical units, 
mapping of original field names to the actual name in the database, etc), 
will be used for catalogue browsing and to process user queries.

The basic user requirements of the proposed database consist of
\item {--} freely accessible via telnet and World Wide Web;
\item {--} inclusion of a sufficiently complete number of source catalogues
 with on-line documentation of their columns;
\item {--} search by position in ($\alpha$, $\delta$)$_{1950}$,
($\alpha$, $\delta$)$_{2000}$ and Galactic coordinates;
\item {--} SQL-type query by user-specified parameters like e.g.~name, 
flux at a given frequency, source size, spectral index, identification;
\item {--} X-window protocols for graphical tools like spectral plots,
radio-sky finding charts and display of two-di\-men\-sional FITS files.

The proposed database will fill a long-standing
gap in multi-waveband astronomical information systems at a time when
several large-area surveys of high sensitivity and resolution 
are in progress with the VLA and the WSRT, producing millions of
further sources. Comparison of available radio data with these new surveys
will be crucial for statistical studies and undoubtedly lead to the
discovery of new source populations.

Cross-identification and spectral classification of many 10,000 radio sources
in our Galaxy and in extragalactic space will provide an unprecedented basis
for radioastronomical and cosmological studies.  Scientific applications of
such a database are extremely numerous, like e.g. the search for 
high-redshift radio galaxies, the study of influence of a cluster environment 
on
radio sources, the selection of targets for future space VLBI mission of
extremely high angular resolution, and the selection of spectrally peculiar
or time-variable sources for follow-up studies.  A much better separation 
of the complex
mix of source populations in the Galactic plane will also be possible.
The project will have its educational impact on students preparing catalogue
documentation and user's guides, and collaborating in specific
research projects.

\References

\ref
Andernach H. 1990, Bull.~Inf.~CDS, 38, 69
\ref
Gubanov A.G. \& Titov V.B. 1996, 
Vestnik St.-Petersburg Univ., Ser. 1, Vyp. 3 (N 15), in press
\ref
Verkhodanov O.V. \& Trushkin S.A. 1995, Preprint 106 SAO RAS, p.~66
\bye